\documentclass[11pt,preprint]{aastex}
\usepackage{pdfpages}
\pdfoutput=1
\begin{document}

\noindent
{\LARGE{\bf The Infinity Pool}}

\vskip 0.3in

\noindent
{\Large Abraham Loeb}

\vskip 0.25in

\noindent
{\Large{\bf Career opportunities are often a matter of chance,
    but also of willingness to cross interdisciplinary boundaries.}}

\vskip 0.35in

\noindent
{\huge R}ecently, my family and I visited Las Casitas village in
Puerto Rico which features an ``infinity pool''. When you enter the
pool, you see the blue ocean behind it meshing smoothly with the
pool's water surface as well as with the blue sky above it (see image
below). The boundaries between these entities are blurred, leaving the
viewer with a sense of freedom to navigate through the continuum of
blended shades of blue. This relaxing view led me to the realization
that much of the tension in our professional life originates from
boundaries, protected by gatekeepers that limit access across them. In
my personal case, I was fortunate to cross the boundary between the
humanities and sciences thanks to the kindness of a few
gatekeepers. The pool of opportunities is infinite, and the talents of
an individual can be realized equally well in completely different
disciplines.  Creative freedom is all about adopting the infinity
pool's point of view, blurring the significance of interdisciplinary
boundaries and continuing to create professionally despite the
``localized'' opinions of some gatekeepers. 

I originally wished to be a philosopher, not a physicist. But
being born in Israel, I was obligated to serve in the military in 1980
when I turned 18. Philosophy appealed to me since it addressed the
most fundamental questions, but I was also good in physics. And so I
was fortunate to be selected for the prestigious Talpiot program which
had been established a year earlier and whose goal was to enable a
group of two dozen recruits per year to pursue intellectual work in
defense-related research in place of the standard military
service. This sounded closer to philosophy than running in the field
as a regular soldier. I felt privileged to have been selected to this
elite group and did my best to justify it. Following three years of
military training and undergraduate studies, we were asked to pursue
work on military or industry projects with immediate practical
applications. But because of my love of philosophy I was driven to
pursue creative intellectual work which was not abundant in these work
environments. I visited a research center which was not on the list of
official work places available to us. Following the visit, I used a
typewriter to produce a formal-looking document outlining a research
project based on some handwritten notes on oil-stained paper that was
handed to me during a lunch meeting by Shalom Eliezer, who later
became my PhD advisor. As a result, this {\it outside of the box}
research proposal was approved at first for a trial period of three
months and later for the remaining five years of my military service
between 1983-1988. I was given some slack by the higher authorities
because I had excelled early on during my military training. Following
a conversation with an experimentalist, Zvi Kaplan, my research
evolved quickly in a new direction, employing a full department, and
was the first project to receive international funding from the
Strategic Defense Initiative (SDI) in the US. The US-based funding
meant that I visited Washington D.C. every few months.

I completed my PhD in plasma physics a couple of years before the end
of my compulsory service and wondered what to do afterwards. One
evening, during a bus ride back from work, I brought up this issue
with a colleague, Arie Zigler, and he mentioned that the most
prestigious postdoc fellowships at that time were awarded by the
Institute for Advanced Study (IAS) at Princeton, where Einstein worked
late in his career. During a subsequent visit to the US, I attended a
conference in Austin, Texas, and met Marshall Rosenbluth who was a
faculty member at IAS two decades earlier. I asked Marshall if he
would recommend that I visit IAS. Marshall's answer was a definite
``Yes''. I immediately called Michelle Sage, the administrative
officer at IAS, and asked if I could visit that coming week. She
replied: ``We do not allow just anyone to visit us. Please send me a
copy of your CV and I will let you know if you can visit''. I mailed
her a list of 11 publications which I had handy and called her again a
few days later. This time she gave me permission to visit, and
scheduled the visit on a Friday at the end of my trip. When I arrived
to her office early in the morning that day, Michelle said: ``There is
one faculty member here with available time, Freeman Dyson. Let me
introduce you to him.'' I was thrilled since I remembered Dyson's name
from textbooks on Quantum Electrodynamics. When I met Freeman in his
office, he said: ``Oh, you are from Israel. Do you know John
Bahcall?'' I said: ``No, I have never heard of him''. Freeman said:
``Let me introduce you to him. John likes Israelis. He is married to
one''.  Fortunately, John was in his office that morning and we had
lunch together. When he heard that Rosenbluth advised me to visit, he
suggested that I visit again for a month in spring 1987, and so I
did. In the meantime John contacted Yuval Ne'eman, the most prominent
Israeli physicist at the time, and asked for more information about
me. Yuval inquired about me and sent a positive report back to
John. At the end of my second visit, John invited me to his office and
said: ``We would like to offer you a five year position if you switch
to astrophysics, but to formally make this offer I need you to arrange
for two recommendation letters''. I was extremely excited and as I ran
down the stairs in E-building which hosted the astrophysicists at IAS,
I saw David Spergel, who had just started the first year of his
postdoc fellowship there. I told David that John had just made me an
offer, and he replied: ``How is that possible?!  The five-year members
are supposed to meet with John this afternoon and discuss the
candidates''. When they did, John asked them: ``Avi looks promising;
should we make him an offer?''. So once again, I had an offer that I
could not decline, even though I really wanted to get back to my old
love of philosophy.

After three years at IAS, I was encouraged to apply to junior faculty
positions, including one at the Harvard Astronomy
department. Eventually Harvard made an offer to another candidate who
declined the position presumably because the prospects for tenure
were slim based on the fact that the previous person who received
tenure from within Harvard Astronomy was Josh Grindlay, a couple of
decades earlier. As a result, I received the offer - which I gladly
accepted because in case of not receiving tenure I could always go
back to my father's farm and work there. After all, I was used to
collecting eggs every afternoon growing up as a child in that farm. I
arrived at Harvard in February 1993.

Three years later, my collaborator Fred Rasio encouraged me to apply
to a faculty position at Cornell university. I did not know anyone at
Cornell but decided to apply since the prospects for tenure at Harvard
were unclear. To my surprise, I received an offer for a tenured
appointment at Cornell, and when I mentioned this offer at Harvard it
was clear that I needed to decide whether to accept or decline the
Cornell offer before Harvard could decide whether to tenure me. I
arranged a meeting with the wisest person on campus, Henry Rosovsky, a
former dean of the Faculty of Arts \& Sciences. Henry asked for some
background information and then advised: ``Stay at Harvard''. I
declined the offer from Cornell and six months later, in December
1996, I received tenure at Harvard.

At that point, it was too late to return to philosophy as my day job,
since I was immersed in an intense research program. Around the same
time, I realized that this arranged marriage was actually to my old
love, dressed up in different clothes. In other words, I figured that
astronomy addresses questions that were previously in the realm of
philosophy or religion, such as: {\it How did the universe start?}
and {\it What is the origin of life?}. Therefore, I actually have the
privilege of addressing philosophical questions using modern
scientific means. In addition, being a theorist rather than an
observer makes me less vulnerable to outside circumstances that are
beyond my control, such as bad weather, allocation of observing time
on telescopes, or long delays in the construction of suitable
instruments. Instead, I can wake up in the morning with an inspiration
for a new idea that was never considered before and flesh it out to a
full paper the same day.

Fifteen years later, the time came to appoint a new Chair to the
Harvard Astronomy department. Another faculty member was offered the
job and declined, so I was offered this job. And three years later my
service was extended to a second three-year term.

The barriers I had to overcome through my unusual career path taught
me freedom in my choice of research topics and diversity in my
selection of collaborators. Interdisciplinary paths often share the
fate of rare seashells swept to the shore which are eroded over time
by ocean waves into indistinguishable sand grains as if they never
existed, unless someone picks them up and preserves them.

Throughout my career, there were several junctions where I could have
been diverted to less fortunate paths. Some of the main opportunities
that benefited me arrived by pure chance. Since things could have
turned out differently, there must be many ``Loebs'' out there with
similar qualifications who did not have these opportunities. With this
in mind, I am deeply committed to help young researchers fulfill their
potential.

About a decade ago, I moved to my current home and discovered a
broken branch on a young tree in the yard. The gardener recommended
cutting off the branch, but close inspection revealed that living
fibers were still linking the branch to the tree and so I chose to tie
the branch to the tree with an insulation tape. Today the branch rises
to the sky far above my height, but the insulation tape is still
visible at my eye level. Every month I stare at the robust branch
whose base swallowed the insulation tape in a slow kiss that lasted
years, and think how important it is to strengthen young researchers
at the fragile beginning of their career.

\bigskip
\bigskip
\bigskip

\noindent{\it Abraham (Avi) Loeb is the Frank B. Baird Jr. Professor
  of Science at Harvard University. He serves as chair of the Harvard
  Astronomy Department and director of the Institute for Theory \&
  Computation at the Harvard-Smithsonian Center for Astrophysics, 60
  Garden Street, Cambridge, Massachusetts 02138, USA. \\e-mail:
  aloeb@cfa.harvard.edu}

\bigskip
\bigskip
\bigskip
\bigskip
\bigskip
\bigskip
\bigskip
\bigskip
\bigskip
\bigskip
\bigskip
\bigskip
\bigskip
\bigskip
\bigskip
\bigskip

\noindent
{\it Image credit:} The infinity pool, Las Casitas Village, Puerto
Rico (A. Loeb, July 5, 2015).

\includepdf[lastpage={1}]{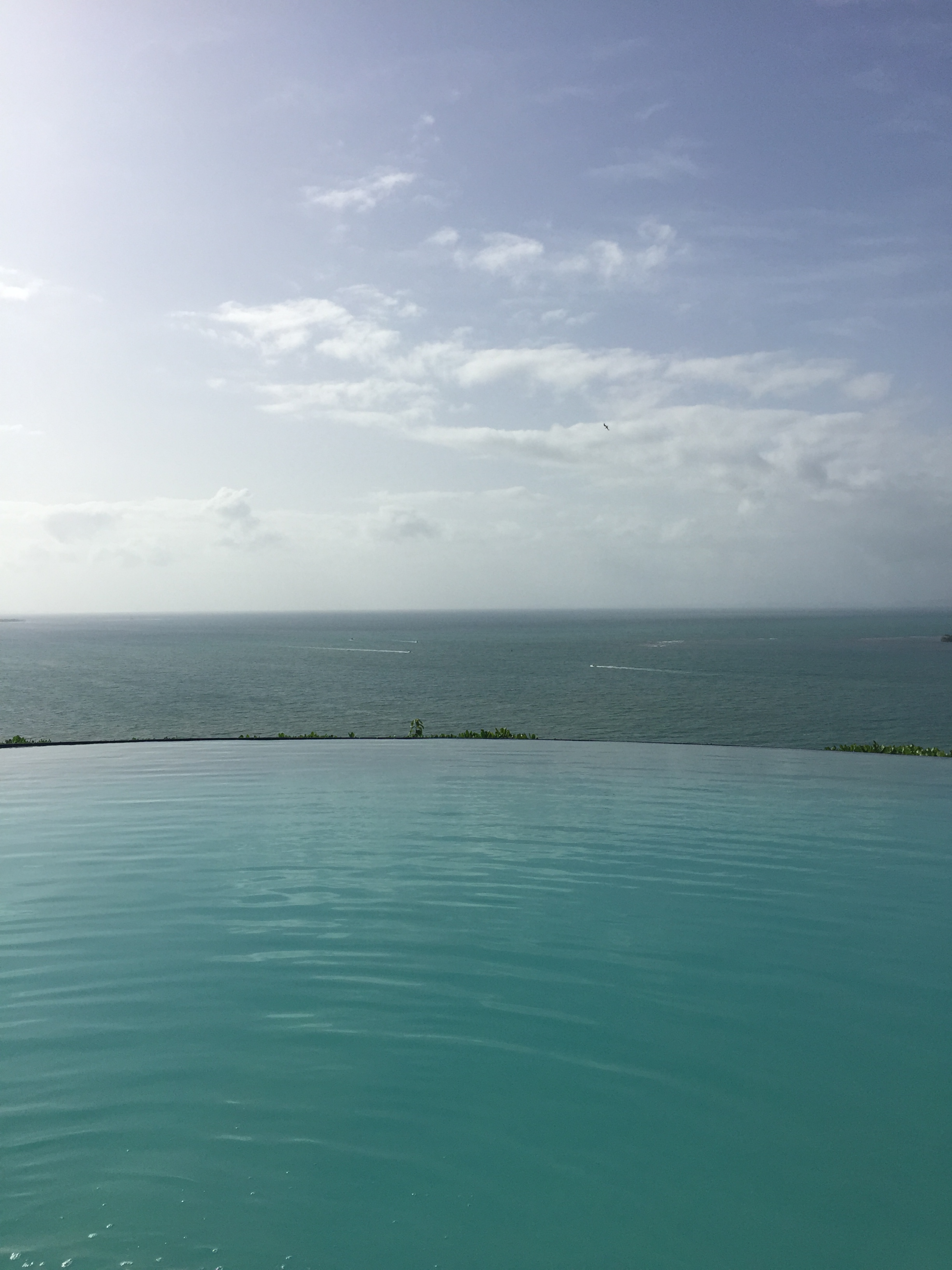}

\end{document}